\documentstyle[11pt,epsfig]{article}

\newlength{\bredde}
\def\slash#1{\settowidth{\bredde}{$#1$}\ifmmode\,\raisebox{.15ex}{/}
\hspace*{-\bredde} #1\else$\,\raisebox{.15ex}{/}\hspace*{-\bredde} #1$\fi}
\newcommand{\mat}{\left ( \begin{array}{cc}}
\newcommand{\emat}{\end{array} \right )}
\newcommand{\matt}{\left ( \begin{array}{ccc}}
\newcommand{\ematt}{\end{array} \right )}
\newcommand{\matf}{\left ( \begin{array}{cccc}}
\newcommand{\ematf}{\end{array} \right )}
\newcommand{\vect}{\left ( \begin{array}{c}}
\newcommand{\evect}{\end{array} \right )}

\newcommand{\be}{\begin{eqnarray}}
\newcommand{\ee}{\end{eqnarray}}
\newcommand{\del}{\partial}
\newcommand{\beq}{\begin{equation}}
\newcommand{\eeq}{\end{equation}}
\newcommand{\ba}{\begin{array}{ccc}}
\newcommand{\ea}{\end{array}}
\newcommand{\nn}{\nonumber}
\newcommand{\noi}{\vspace{6pt}\noindent}
\newcommand{\lG}{\raise.3ex\hbox{$\stackrel{\leftarrow}{G}$}}
\newcommand{\lU}{\raise.3ex\hbox{$\stackrel{\leftarrow}{U}$}}
\newcommand{\lP}{\raise.3ex\hbox{$\stackrel{\leftarrow}{{\cal P}}$}}
\newcommand{\leta}{\raise.3ex\hbox{$\stackrel{\leftarrow}{\eta}$}}
\newcommand{\lOmega}{\raise.3ex\hbox{$\stackrel{\leftarrow}{\Omega}$}}
\newcommand{\ldr}{\raise.3ex\hbox{$\stackrel{\leftarrow}{\delta^r}$}}

\def\beqn{\begin{eqnarray}}
\def\eeqn{\end{eqnarray}}

\def\gtwid{\raise.3ex\hbox{$>$\kern-.75em\lower1ex\hbox{$\sim$}}}
\def\ltwid{\raise.3ex\hbox{$<$\kern-.75em\lower1ex\hbox{$\sim$}}}

\def\matA{{\mbox{\bf A}}}
\def\matB{{\mbox{\bf B}}}

\def\matD{{\mbox{\bf D}}}

\begin{document}
\topmargin -1.4cm
\oddsidemargin -0.8cm
\evensidemargin -0.8cm

\title{\Large{{\bf The Replica Method and Toda Lattice Equations for QCD$_3$}}}

\vspace{1.5cm}

\author{~\\
{\sc T. Andersson$^a$, P.H. Damgaard$^a$} and {\sc K. Splittorff$^b$}\\
$^a$The Niels Bohr Institute and $^b$NORDITA\\ Blegdamsvej 17\\ 
DK-2100 Copenhagen {\O}\\
Denmark
}
\maketitle
\vfill
\begin{abstract}
We consider the $\epsilon$-regime of QCD in 3 dimensions. It is shown that
the leading term of the effective partition function satisfies a set of 
Toda lattice equations, recursive in the number of flavors. Taking the 
replica limit of these Toda equations allows us to derive the
microscopic spectral correlation functions for the QCD Dirac operator in 3
dimensions. For an even number of flavors we reproduce known results derived
using other techniques. In the case of an odd number of flavors the theory
has a severe sign problem, and we obtain previously unknown microscopic
spectral correlation functions.    
\end{abstract}
\vfill


\begin{flushright}
NORDITA-2004-79 HE 
\end{flushright}

\thispagestyle{empty}
\newpage

\setcounter{equation}{0}
\section{Introduction}
\vspace*{0.3cm}

\noi
Recently there has been substantial progress in understanding the
so-called replica method in the context of the effective low-energy 
field theory for QCD (QCD$_4$) [1-7].
While perturbative (series) expansions readily
yield correct results 
with this method \cite{DS}, it has proved difficult to obtain exact 
nonperturbative results \cite{critique,KM,lerner,zirn}.
A crucial ingredient in formulating a systematic
replica method which is reliable also for nonperturbative calculations 
has been the use of exact relations of the  Painlev\'e \cite{K} or Toda
\cite{SplitVerb1,SplitVerb2,SplitVerb3} kinds.
These relations reflect the fact \cite{GLeps} that the leading-order effective
partition function in the so-called $\epsilon$-regime \cite{GLeps}\footnote{
See $e.g.$ ref. \cite{Dreview} for a recent review with an emphasis on
the present context.} of QCD is what is known as a $\tau$-function of
an underlying integrable KP hierarchy \cite{AD}. This integrable structure 
derives back
to the spontaneous breaking of chiral symmetry according to the
pattern SU$_L(N_f)\times$SU$_R(N_f) \to$ SU($N_f$), where $N_f$ is the
number of light quark flavors. As a consequence of this spontaneous breaking
of chiral symmetry and because of the mass gap in QCD the low-energy dynamics
of QCD is governed by the Goldstone modes. In the standard counting scheme
the low energy theory describing the dynamics of the Goldstone modes 
is known as chiral perturbation theory, and in this
framework the ordinary replica method applies without subtleties \cite{DS}. 
If one instead \cite{GLeps} 
considers a counting scheme (the $\epsilon$-regime) where the Compton 
wavelength of the pions is much
larger than the linear dimension of the volume, $1/m_\pi \gg L$, then the
partition function at leading order reduces to a static integral over the
Goldstone manifold. Changing $N_f$ in this integral builds up a series 
of partition functions which
are connected through a Toda lattice equation. This connection between
partition functions with different $N_f$ shows how to take the replica
limit $N_f\to0$.
The purpose of this paper is to establish Toda 
lattice equations for QCD in 3 dimensions (QCD$_3$) and to show that also
these allow us to obtain exact non-perturbative results using the replica
method.  

\noi
QCD$_3$ cannot
undergo spontaneous breaking of chiral symmetry in the usual sense.
Moreover, also the notion of gauge field topology is
different, a remnant manifestation being the possibility
of adding a Chern-Simons term to the action. Nevertheless, a
three-dimensional analogue of chiral symmetry breaking (really
flavor symmetry breaking, re-interpreted) is possible \cite{P}.
The basic idea is that the chiral components of
the four-dimensional spinors in three dimensions can be mimicked
by two different fermion fields, with masses of equal magnitude
but opposite signs. Such mass terms of different signs are possible
in an odd number of dimensions, a consequence of the fact the two sets 
of $\gamma$-matrices $\{\gamma_i\}$ and $\{-\gamma_i\}$ form two
inequivalent irreducible representations of the Clifford algebra
in an odd number of space-time dimensions.
The two 2-spinors corresponding to opposite signs of the mass terms
may be grouped into one four-spinor of which the top components
play the r\^{o}le of the left-handed field, while the lower
components represent the analogue of the right-handed field in
four-dimensional language. Moreover, the associated ``chiral''
symmetry can break spontaneously, and is {\em expected} to break
spontaneously when the number of fermions species is small. 
The symmetry breaking pattern is believed to be that of \cite{P,VZ}
\beq
U(2N_f) \to U(N_f)\times U(N_f) . \label{evenpattern}
\eeq
There are numerical simulations of QCD$_3$ on rather small lattices
that support this conclusion in the
extreme (quenched) case of no dynamical quarks \cite{DHKM}.  
An odd number of flavors is obviously a more difficult issue. For
$2N_f+1$ flavors of which the $2N_f$ masses are grouped into pairs
of equal magnitude but opposite signs 
it has been argued in ref. \cite{VZ} that the spontaneous
symmetry breaking pattern (\ref{evenpattern})
is replaced by 
\beq
U(2N_f+1) \to U(N_f+1) \times U(N_f) ~. 
\eeq
This case is rather tricky to treat from the effective field theory
point of view. An interesting starting point is Random Matrix Theory,
which suggests that there are two possible effective field theories,
depending on the number of Dirac operator eigenvalues \cite{VZ}. We 
will look into this problem in closer detail below. We note that as
one further generalization one could also consider the case of $2N_f$
flavors paired with opposite signs, and $n$ unpaired flavors.

\noi
The Random Matrix Theory representation of the effective partition 
function for QCD$_3$ in the $\epsilon$-regime \cite{VZ} has lead to 
a number of intriguing conjectures.
In particular, spectral properties of the pertinent 
3-dimensional Dirac operator have been argued to be derivable from the joint 
eigenvalue probability distributions of such a Random Matrix Theory 
\cite{VZ,Nagao}, 
and based on this conjecture universal spectral
$n$-point correlation functions have been found 
for both an even \cite{ADMN,DN,AD1} and an odd \cite{C} number of flavors. 
First steps toward deriving these results directly from the effective
field theory partition function were taken in refs. \cite{S,ADDV}. 
In particular, by an impressive series of supermanifold integrations some 
of the spectral correlation functions were derived from the so-called
``supersymmetric'' formulation in ref. \cite{S}.
In \cite{ADDV} intriguing relations between the QCD$_4$ and the QCD$_3$
partition functions in the $\epsilon$-regime were derived. Applying the 
replica method to these relations a set of identities between the
spectral correlation followed. These identities, derived by formally 
applying the replica limit $N_f \to 0$, presupposed a meaningful operational
way to perform this limit, a non-trivial issue in view of the known
subtleties involved \cite{critique,zirn}.
Here, using the replica limit of the Toda lattice equation, 
we show that these results, and others which are new, can be 
given a precise meaning.

\noi
The new results are for an odd number of flavors. This theory is rather
special since after integrating out the fermions
the weight inside the partition function is
not positive definite. This manifest itself in the eigenvalue density in a
number of ways. For example, it is tempting to conclude that eigenvalue
density in the theory with and odd number of flavors and an odd
number of Dirac operator eigenvalues
\cite{VZ} is an odd function in the chiral limit. However, 
as we will explain below this is not the case. 
Although the non-positive weight in
QCD$_3$ with an odd number of flavors comes about quite differently from the
sign problem in QCD$_4$ with a non-zero baryon chemical potential, the
lessons learned from QCD$_3$ may be worth keeping in mind when tackling
the sign problem in QCD$_4$.

\noi
The presentation of this paper is as follows. First we derive the Toda
lattice equations for QCD$_3$ in the $\epsilon$-regime. Then we discuss the 
extension to bosonic flavors. Given the Toda lattice equations we take the
replica limit in section \ref{sec:replicalim} and show how to
obtain spectral
correlation functions. As in QCD$_4$, the Toda lattice equations in QCD$_3$ 
also hold for partition functions with both fermions and bosons. This allows
us to establish the replica limit of the Toda lattice equation in QCD$_3$ 
on the same footing as the supersymmetric method \cite{Efetov}.
We show how in section \ref{sec:susy}.
Finally, in section \ref{sec:consist} we show how
general Toda lattice equations follow from one single Consistency 
Condition of Random Matrix Theory. We sumarize our results in
section 6.

\setcounter{equation}{0}
\section{The Toda Lattice: From QCD$_4$ to QCD$_3$}
\vspace*{0.3cm}

Our starting point is a set of Toda lattice equations which have
been derived \cite{Kharchev,Witte,AD,SplitVerb2,SplitVerb3} for the 
leading term of the effective 
QCD$_4$ partition function in the $\epsilon$-regime. Using these and known
connections between QCD$_4$ and QCD$_3$ in the $\epsilon$-regime
we establish the Toda lattice equations for QCD$_3$. 

\noi
We first give our
conventions and definitions, which follow ref. \cite{ADDV}. 
The $\epsilon$-regime is in this
(2+1)-dimensional context defined by the large-volume limit
where nevertheless $V \ll 1/m_\pi^3$ and $m_{\pi}$ generically
denotes the pseudo-Goldstone masses of spontaneous breaking of chiral 
symmetry. In this ``extreme'' chiral limit where $\mu_i \equiv m_i V \Sigma$
is kept fixed the partition function reduces to a static integral over the
Goldstone manifold. The infinite-volume condensate is defined by 
\beq
\Sigma ~\equiv~ 
\lim_{m\to 0}\lim_{V\to\infty}|\langle\bar{\psi}\psi\rangle| ~,
\eeq
and $m_i$ are the quark masses.

\noi
The leading contribution to the QCD$_3$ partition function in the
$\epsilon$-regime is for an even number $2N_f$ of fermions grouped in
pairs with masses of equal magnitude but opposite signs given by
\cite{VZ}
\beq
{\cal Z}_{3}^{(2N_f)}(\{\mu_i\})\ =\ \int_{U(2N_{f})}\! dU \exp
[V\Sigma {\mbox{\rm Re~Tr}({\cal M}}U\Gamma_5 U^\dagger)]\ ,
\label{ZQCD3even}
\eeq
with  ${\cal M}$=diag$(m_1,\ldots,m_{N_f},-m_1,\ldots,-m_{N_f})$,
$\Gamma_5 \equiv$ diag({\bf 1}$_{N_f}$,-{\bf 1}$_{N_f}$) and $dU$
always indicating the Haar measure. 
This integral can be performed explicitly with the help of the
Itzykson-Zuber formula \cite{IZ}, resulting in \cite{DN}
\beq
{\cal Z}^{(2N_{f})}_{3}(\{\mu_i\}) ~=~ (-1)^{N_f(N_f+1)/2}
\ \frac{\det\left(
\begin{array}{ll}
\matA(\{\mu_i\}) & \matA(\{-\mu_i\})\\
\matA(\{-\mu_i\})  & \matA(\{\mu_i\})
\end{array}\right)}{\Delta({\cal M})} ~,\label{ZQCD3evenexpl}
\eeq
where, with our normalization conventions, the $N_f\!\times\!N_f$ matrix
$\matA(\{\mu_i\})$ is defined by
\beq
\matA(\{\mu_i\})_{jl} ~\equiv~ (\mu_j)^{l-1} \mbox{e}^{\mu_j} 
~~,~~~~j,l=1,\ldots,N_f~,
\label{matAdef}
\eeq
and the Vandermonde determinant is given by
\beq
\Delta({\cal M}) ~=~ \prod_{i>j}^{2N_{f}}(\mu_i-\mu_j) ~.
\eeq

\noi
As mentioned in the Introduction, the case of an odd number of
flavors is a bit unusual because the symmetry of the action 
relies on a paring of masses.  
Nevertheless, as in the even-flavored theory, the effective partition
function has a unique leading-order term in the $\epsilon$-scheme. With ${\cal
  M}$=diag$(\{\mu_i\},\mu,\{-\mu_i\})$ and  
$\tilde \Gamma_5 = ({\bf 1}_{N_f+1},-{\bf 1}_{N_f})$ the partition 
function becomes \cite{VZ}, 
\be
{\cal Z}_{3^+}^{(2N_f+1)}(\mu,\{\mu_i\})\ 
=\ \int_{U(2N_{f}+1)}\! dU \cosh
[V\Sigma {\mbox{\rm Re~Tr}({\cal M}}U \tilde\Gamma_5 U^\dagger)]\ ,
\label{ZQCD3odd0}
\ee
for an even number of Dirac operator eigenvalues, and
\be
{\cal Z}_{3^-}^{(2N_f+1)}(\mu,\{\mu_i\})\ 
=\ \int_{U(2N_{f}+1)}\! dU \sinh
[V\Sigma {\mbox{\rm Re~Tr}({\cal M}}U\tilde \Gamma_5 U^\dagger)]\ ,
\label{ZQCD3odd1}
\ee
for an odd number of Dirac operator eigenvalues. Both of these
integrals can again be evaluated explicitly, with the conventions
of \cite{ADDV},
\beq
{\cal Z}^{(2N_{f}+1)}_{3^\pm}(\mu,\{\mu_i\}) ~=~ (-1)^{N_f(N_f+3)/2}
\frac{2^{N_f}}{\Delta({\cal M})}\frac12 
\left[ \det \matD(\mu,\{\mu_i\}) \pm (-1)^{N_f}
\det \matD(-\mu,\{-\mu_i\})\right]~,
\label{ZQCD3oddexpl}
\eeq  
where the $(2N_f+1)\times(2N_f+1)$ matrix $\matD$ is defined as
\beq
2^{N_f} \det \matD(\mu,\{\mu_i\}) ~\equiv~
\det\left(
\begin{array}{ll}
\matA(\mu,\{\mu_i\})_{N_f+1\times N_f+1} 
& \matA(-\mu,\{-\mu_i\})_{N_f+1\times N_f}\\
\matA(\{-\mu_i\})_{N_f\times N_f+1}  
& \matA(\{\mu_i\})_{N_f\times N_f}
\end{array}\right) ~,
\label{matDdefodd}
\eeq
and $\matA$ is defined in eq. (\ref{matAdef}).

\noi
For QCD$_4$ in the corresponding $\epsilon$-regime the leading contribution
to the partition function is\footnote{We use a notation in which we do
not explicitly indicate that this is the 4-dimensional partition function;
the partition function is seen to be the one for QCD$_4$ by the
labeling according to topological charge $\nu$. There should be no
source of confusion as we will never explicitly consider the case $\nu=3$.}
\beq
{\cal Z}_{\nu}^{(N_f)}(\{\mu_i\}) ~=~
\int_{U(N_f)}\! dU~(\det U)^{\nu}\exp\left[V\Sigma{\rm Re Tr}({\cal M}U)
\right]
\eeq
in a sector of topological charge $\nu$ \cite{GLeps,LS}. The quark mass 
matrix in 4 
dimensions is ${\cal M}$=diag$(m_1,\ldots,m_{N_f})$. Also this integral
can be explicitly evaluated for integer $\nu$ \cite{Andy}: 
\beq
{\cal Z}_\nu^{(N_f)} (\{\mu_i\}) ~=~ \frac{\det \matB(\{\mu_i\})}
{\Delta(\{\mu_i^2\})},
\label{ZchUEexpl}
\eeq
where the matrix $\matB$ in eq. (\ref{ZchUEexpl}) is given by
\beq
\matB(\{\mu_i\})_{jl} 
= \mu_j^{l-1}I_{\nu}^{(l-1)}(\mu_j)~~,~~~~j,l=1,\ldots,N_f ~,
\label{matBdef}
\eeq
with $I_{\nu}^{(l)}$ the $l$'th derivative modified Bessel function 
$I_{\nu}$, and the denominator is given by the Vandermonde determinant of, 
in this case, squared quark masses,
\beq
\Delta (\{\mu_i^2\})\ \equiv\ \prod_{i>j}^{N_f}(\mu_i^2-\mu_j^2)
\ =\ \det_{i,j}\left[ (\mu_i^2)^{j-1}\right] \ . 
\label{Vandermonde}
\eeq
We use on purpose the same notation $\mu_i$ in both
QCD$_3$ and QCD$_4$, since relations between these two widely
different theories exist when these dimensionless parameters are identified
in the two theories. The QCD$_4$ partition function for non-integer
$\nu$ is defined by analytical continuation in the index $\nu$ of the
Bessel functions \cite{ADDV}. We stress at this point that the explicit
formulas (\ref{ZQCD3evenexpl}), (\ref{ZQCD3oddexpl}) and 
(\ref{ZchUEexpl}) of course are only valid for $N_f$ taking positive integer 
values. Below
we shall return to the issue of how 
to deal with cases where $N_f$ is zero, or even negative.

\subsection{Two theorems}

Our main results will be based on the following two relations between the
QCD$_3$ and QCD$_4$ partition functions. These relations were 
proven in ref. \cite{ADDV} and hold when the
normalization conventions are as stated above.

\noi
{\sc Theorem I} 

\noi
\beq
{\cal Z}_{3}^{(2N_f)}(\{\mu_i\}) ~=~ 
\pi^{N_f}{\cal Z}_{\nu=-1/2}^{(N_f)}(\{\mu_i\})
                   ~ {\cal Z}_{\nu=+1/2}^{(N_f)}(\{\mu_i\}) ~.
\label{factoreven}
\eeq

\vspace{0.5cm}
\noi\noindent
{\sc Theorem II} 

\noi
\beq
{\cal Z}_{3^\pm}^{(2N_f+1)}(\mu,\{\mu_i\}) ~=~ 
\pi^{N_f}
\sqrt{\frac{\pi\mu}{2}}\ {\cal Z}_{\nu=\mp 1/2}^{(N_f+1)}(\mu,\{\mu_i\}) 
            \ {\cal Z}_{\nu=\pm 1/2}^{(N_f)}(\{\mu_i\}) \ .
\label{factorodd}
\eeq
We have written these theorems explicitly for real and positive
masses $\{\mu_i\}$. When taking discontinuities in the complex plane
one should always recall that only the absolute value enters
here. In what follows we will for notational simplicity
normally not explicitly display the
associated factors that equal unity for real and positive values
(usually terms of the form $x/\sqrt{x^2}$).

\subsection{Toda lattice equations for QCD$_3$}
\label{sec:TodaQCD3}

The two theorems (\ref{factoreven}) and (\ref{factorodd})
can nicely be combined to derive Toda lattice
equations for QCD$_3$ on the basis of the known Toda equations for
QCD$_4$. As a first illustration of this procedure,
consider the case of $N_f$ degenerate flavors of mass $x$ in QCD$_4$. In
this case the leading term of the QCD$_4$ partition function
satisfies the differential equation \cite{Kharchev}
\beq
(x\partial_x)^2\ln{\cal Z}^{(N_f)}_{\nu}(x) ~=~
2N_fx^2\frac{{\cal Z}^{(N_f+1)}_{\nu}(x){\cal Z}^{(N_f-1)}_{\nu}(x)}
{\left[{\cal Z}^{(N_f)}_{\nu}(x)\right]^2} ~. \label{toda4}
\eeq
We can turn this into a Toda equation for QCD$_3$ by use of
Theorem I and II. By considering the case of an even number
of flavors $2N_f$ with degenerate masses $x$ we get
\beq
(x\partial_x)^2\ln{\cal Z}^{(2N_f)}_3(x)
= 4N_fx\frac{{\cal Z}^{(2N_f+1)}_{3^+}(x){\cal Z}^{(2N_f-1)}_{3^-}(x)
+ {\cal Z}^{(2N_f+1)}_{3^-}(x){\cal Z}^{(2N_f-1)}_{3^+}(x)}
{\left[{\cal Z}^{(2N_f)}_{3}(x)\right]^2} ~. \label{toda3}
\eeq
This is the QCD$_3$ Toda lattice equation analogous to the
QCD$_4$ equation (\ref{toda4}). We note the quite generic feature of the
QCD$_3$ equation having two contributions on the right hand side,
originating from having two different partition functions with
an odd number of flavors. Of course, we might have derived the
Toda equation (\ref{toda3}) directly from the explicit formulas
(\ref{ZQCD3evenexpl}) and (\ref{ZQCD3oddexpl}). But the present
shortcut through use of Theorems I and II is obviously a much simpler
route.

\noi
We next consider the QCD$_4$ Toda lattice equations with 
$N_f$ degenerate flavors of mass $x$ and $n$ degenerate flavors of mass
$y$. These are \cite{Kharchev}
\beq
x\partial_x(x\partial_x +y\partial_y)\ln{\cal Z}^{(N_f,n)}_{\nu}(x,y) ~=~
2N_fx^2\frac{{\cal Z}^{(N_f+1,n)}_{\nu}(x,y){\cal Z}^{(N_f-1,n)}_{\nu}(x,y)}
{\left[{\cal Z}^{(N_f,n)}_{\nu}(x,y)\right]^2} ~, \label{toda42}
\eeq
and \cite{SplitVerb2}
\beq
(x\partial_x y\partial_y)\ln{\cal Z}^{(N_f,n)}_{\nu}(x,y) ~=~
4N_fnx^2 y^2\frac{{\cal Z}^{(N_f+1,n+1)}_{\nu}(x,y)
{\cal Z}^{(N_f-1,n-1)}_{\nu}(x,y)}
{\left[{\cal Z}^{(N_f,n)}_{\nu}(x,y)\right]^2} ~. \label{toda43}
\eeq

\noi
The corresponding Toda lattice equations for the QCD$_3$ partition functions 
are readily obtained from the theorems. To keep it simple, we will start 
by looking at $2N_f$ flavors with degenerate 
masses $\pm x$ and $2n$ flavors with degenerate masses $\pm y$ (both with the
usual pairing). After re-expressing the Toda equations in terms
of QCD$_3$ partition functions we get

\be
&& x\partial_x (x\partial_x + y\partial_y) 
\ln {\cal Z}^{(2N_f,2n)}_{3}(x,y) \nonumber \\ 
&=& 4 N_f x \frac{{\cal Z}^{(2N_f+1,2n)}_{3^+}(x,y) 
{\cal Z}^{(2N_f-1,2n)}_{3^-}(x,y) + {\cal Z}^{(2N_f+1,2n)}_{3^-}(x,y) 
{\cal Z}^{(2N_f-1,2n)}_{3^+}(x,y)}{\left[{\cal Z}^{(2N_f,2n)}_{3}(x,y)\right]^2
} ~. 
\label{toda32}
\ee 
Once again the right hand side has two contributions. The generalization 
to the case where the $2n$ masses are paired but non-degenerate is 
obtained simply by replacing $y$ with $\{y_i\}$ and 
$(x\partial_x+y\partial_y)$ by 
$(x\partial_x + \sum_{i=1}^n y_i \partial_{y_i})$.
Following the same route as described above, we find that
the QCD$_3$ equivalent of (\ref{toda43}) for an even number of flavors is
\be
&& x\del_x y\del_y \log {\cal Z}^{(2N_f,2n)}_{3}(x,y) \\
&=& 4N_{f} n x^3 y \left(\frac{{\cal Z}_{3^-}^{(2N_{f}+2,2n+1)}(x,y)
{\cal Z}_{3^+}^{(2N_{f}-2,2n-1)}(x,y)}
{{\cal Z}_{3^+}^{(2N_f+1,2n)}(x,y){\cal Z}_{3^-}^{(2N_f-1,2n)}(x,y)}
+\frac{{\cal Z}_{3^+}^{(2N_{f}+2,2n+1)}(x,y)
{\cal Z}_{3^-}^{(2N_{f}-2,2n-1)}(x,y)}
{{\cal Z}_{3^-}^{(2N_f+1,2n)}(x,y)
{\cal Z}_{3^+}^{(2N_f-1,2n)}(x,y)} \right) ~.\nn
\label{toda33} 
\ee 
Despite appearences, the right hand side is symmetric under interchange of 
$x$ and $y$.


\noi
We now turn to the partition functions with an odd number of 
flavors. Starting at the simplest case with $2N_f$ flavors of paired
masses $\pm x$ and one additional flavor of mass $x$ we get from (\ref{toda4}):
\be
&& (x\partial_x)^2 \ln {\cal Z}^{(2N_f+1)}_{3^\pm}(x) \\
&=& 4 x \left( (N_f+1) \frac{{\cal Z}^{(2N_f+3)}_{3^{\pm}}(x)
{\cal Z}^{(2N_f+1)}_{3^{\mp}}(x)}
{\left[{\cal Z}^{(2N_f+2)}_{3}(x)\right]^2}+ 
N_f \frac{{\cal Z}^{(2N_f+1)}_{3^{\mp}}(x)
{\cal Z}^{(2N_f-1)}_{3^{\pm}}(x)}
{\left[{\cal Z}^{(2N_f)}_{3}(x)\right]^2} \right) ~.\nonumber 
\label{toda31unev}
\ee 

\noi
For the QCD$_3$ version of (\ref{toda42}) where we have $2N_f$ paired and
degenerate masses $\pm x$ and $2n+1$ flavors of mass $y$ we find
\be
&& x\del_x (x\del_x + y\del_y) \ln {\cal
Z}^{(2N_f,2n+1)}_{3^\pm}(x,y) \\ 
&=& 4 N_f x \left( \frac{{\cal Z}^{(2N_f+1,2n+2)}_{3^\pm}(x,y) 
{\cal Z}^{(2N_f-1,2n+2)}_{3^\mp}(x,y)}
{\left[{\cal Z}^{(2N_f,2n+2)}_{3}(x,y)\right]^2} 
+ \frac{{\cal Z}^{(2N_f+1,2n)}_{3^\mp}(x,y){\cal Z}^{(2N_f-1,2n)}_{3^\pm}(x,y)}
{\left[{\cal Z}^{(2N_f,2n)}_{3}(x,y)\right]^2}\right) ~.\nonumber
\label{toda32unev}
\ee
As the final Toda lattice equation for QCD$_3$ we also 
give the analogue of (\ref{toda33}), 
\be
&& x\del_x y\del_y \log {\cal Z}^{(2N_f,2n+1)}_{3^{\pm}}(x,y) \\
&=& 4N_{f} x^3 y \left((n+1)\frac{{\cal Z}_{3^{\pm}}^{(2N_{f}+2,2n+3)}(x,y)
{\cal Z}_{3^{\mp}}^{(2N_{f}-2,2n+1)}(x,y)}
{{\cal Z}_{3^\mp}^{(2N_f+1,2n+2)}(x,y){\cal Z}_{3^\pm}^{(2N_f-1,2n+2)}(x,y)}
\right.\nonumber \\ 
&& \hspace{2cm} \left. 
+n\frac{{\cal Z}_{3^\mp}^{(2N_{f}+2,2n+1)}(x,y)
{\cal Z}_{3^\pm}^{(2N_{f}-2,2n-1)}(x,y)}
{{\cal Z}_{3^\pm}^{(2N_f+1,2n)}(x,y){\cal
Z}_{3^\mp}^{(2N_f-1,2n)}(x,y)} \right) ~,  \nonumber 
\label{toda33unev}
\ee
valid for an odd number of flavors.

\subsection{Extension to bosonic flavors}

Before proceeding to use the replica method we must ascertain that
the Theorems I-II are consistent with what has been understood
about the QCD$_4$ partition functions for both zero and a negative
number of flavors. We should first clarify what is meant by this.
For zero flavors the partition functions are ``quenched'', the
fermion determinants are entirely absent, and there can be no
mass dependence. It is clearly convenient to choose the normalization
so that in this case the partition function simply equals unity.
A negative number of fermionic flavors is defined by raising
the determinant to the corresponding negative number. This is
equivalent to a partition function of the same number of complex
fields with bosonic statistics (and thus violating the 
spin-statistics theorem, but these bosons are never considered
as external physical states). On the QCD$_4$ side the generalization of the 
partition function formula (\ref{ZchUEexpl}) was given in
ref. \cite{FSgen,SplitVerb1,FA} 
for an arbitrary number $N_f$ and $N_b$ of fermionic and bosonic species,
respectively. For {\em positive} $N_f$ and $N_b$ the explicit
expression reads, in a hopefully obvious notation,
\be
\label{Z-nm}
{\cal Z}_\nu^{(N_f|N_b)}(\{x_f\}|\{y_b\})\!=\!\frac{\det[z^{j-1}_i{\cal
J}_{\nu+j-1}(z_i)]_{i,j=1,..,N_f+N_b}}
{\prod_{j>i=1}^{N_f}(x_j^2-x_i^2)\prod_{j>i=1}^{N_b}(y_j^2-y_i^2)},
\ee
where $z_i=x_i$ for $i=1,\ldots,N_f$,\, $z_{N_f+i}=y_i$ for
$i=1,\ldots,N_b$, \,${\cal J}_{\nu+j-1}(z_i)\equiv I_{\nu+j-1}(x_i)$ for
$i=1,\ldots,N_f$, and ${\cal J}_{\nu+j-1}(z_{N_f+i})\equiv
(-1)^{j-1}K_{\nu+j-1}(y_i)$ for $i=1,\ldots,N_b$.
The generalization to negative integers is done through identifications
such as ${\cal Z}_{\nu}^{(-N_f|N_b)} = {\cal Z}_{\nu}^{(0|N_f+N_b)}$
and so on.
Thus, while the $N_f=1$ QCD$_4$ partition function is given by
${\cal Z}^{(1)}_{\nu}(x)=I_{\nu}(x)$, the $N_f=-1$ partition function
reads ${\cal Z}^{(-1)}_{\nu}(x)=K_{\nu}(x)$, where $K_{\nu}(x)$ is the
modified Bessel function. We therefore know how to define the right
hand side of the two theorems for a negative number
of flavors. The question is whether we can define the left hand side
for zero or a negative number of flavors in QCD$_3$ in this way.
Setting $N_f=0$
in eq. (\ref{factoreven}) gives ${\cal Z}_3^{(0)} = 1$ from Theorem I, 
$i.e$, with
the correct normalization. Having gained some faith in this procedure, we
can infer the QCD$_3$ partition function for $N_f=-1$ from
Theorem II:
\beq
{\cal Z}_{3^{\pm}}^{(-1)}(x) ~=~ \sqrt{\frac{x}{2\pi}}K_{\pm 1/2}(x) 
~=~ e^{-x} ~.
\eeq
This agrees exactly with what we should expect on general principle:
For $N_f=-1$ (one boson) there is no spontaneous symmetry breaking
at all, and the leading term of the partition function is simply, by
a generalization of the argument by Leutwyler and Smilga in the
4-dimensional case \cite{LS}, the exponentiation of the leading
term free energy $F= mV\Sigma = \mu$. It thus appears that
we can continue with
the indicated identifications, and
Theorem I will then give us the partition function for $N_f=-2$:
\beq
{\cal Z}_3^{(-2)}(x) ~=~ \frac{1}{\pi}K_{-1/2}(x)K_{1/2}(x) ~.
\eeq
This procedure obviously continues for higher (negative) values of
$N_f$. 

\noi
We can formalize the above arguments by making the following
conjectures regarding supersymmetric generalizations of
Theorems I and II:
\be
{\cal Z}_3^{(2N_f|2N_b)}(\{x_i\}|\{y_i\}) &=& \pi^{N_f-N_b}
{\cal Z}_{-1/2}^{(N_f|N_b)}(\{x_i\}|\{y_i\})
{\cal Z}_{1/2}^{(N_f|N_b)}(\{x_i\}|\{y_i\}) \\
{\cal Z}_{3^{\pm}}^{(2N_f+1|2N_b)}(\{x_i\},x|\{y_i\}) &=& \pi^{N_f-N_b}
\sqrt{\frac{\pi x}{2}}
{\cal Z}_{\mp 1/2}^{(N_f+1|N_b)}(\{x_i\},x|\{y_i\})
{\cal Z}_{\pm 1/2}^{(N_f|N_b)}(\{x_i\}|\{y_i\}) \\
{\cal Z}_{3^{\pm}}^{(2N_f|2N_b+1)}(\{x_i\}|\{y_i\},y) &=& \pi^{N_f-N_b}
\sqrt{\frac{y}{2\pi}}
{\cal Z}_{\pm 1/2}^{(N_f|N_b+1)}(\{x_i\}|\{y_i\},y)
{\cal Z}_{\mp 1/2}^{(N_f|N_b)}(\{x_i\}|\{y_i\}) ~.\label{bosth2}
\ee
In all cases we have been able to check these identities are
consistent with what we know from other sources. We note in particular
that if they are correct we have circumvented the quite
difficult task of evaluating all the so-called Efetov-Wegner terms
in the supersymmetric version of the effective Lagrangian \cite{S}.
It appears that the most direct way to prove these identities may
be through an explicit evaluation of the related supersymmetric
Random Matrix Theory integral, as in ref. \cite{FA}.
%

\vspace{0.5cm}
\section{Replica limit of the QCD$_3$ Toda lattice equations}
\label{sec:replicalim}
\vspace*{0.3cm}

As an introduction let us consider QCD$_4$ with $N_f$ quarks with masses
$\{\mu_f\}$ and $n$ fermions of mass $x$. Using the replica method the 
partially quenched chiral condensate, the resolvent, is then defined as 
\beq
G_\nu^{(N_f)}(x,\{\mu_f\})\equiv\lim_{n\rightarrow 0}
\frac{1}{n}\frac{\partial}{\partial x}
\ln {\cal Z}_{\nu}^{(N_f,n)}(\{\mu_f\},x) ~.
\label{res4}
\eeq 
The spectral density of the Dirac operator in QCD$_4$ with $N_f$
flavors is given by 
\be
\rho_\nu^{(N_f)}(\lambda,\{\mu_f\}) \equiv
\left\langle\sum_{k}\delta(\lambda_k-\lambda)\right\rangle_{N_f} ,
\label{rhodef}
\ee
where $\langle\ldots\rangle_{N_f}$ is the vacuum expectation value in QCD$_4$
with $N_f$ flavors.
The delta functions can be expressed as the discontinuity of the 
resolvent across the imaginary axis
\be
\rho_\nu^{(N_f)}(\lambda,\{\mu_f\}) &=& \frac{1}{2\pi} {\mbox{\rm Disc}}
\left.G_\nu^{(N_f)}(x,\{\mu_f\})\right|_{x=i\lambda} \nonumber \\
&=& \frac{1}{2\pi}\lim_{\varepsilon\rightarrow 0}
[G_\nu^{(N_f)}(i\lambda+\varepsilon) - G_\nu^{(N_f)}(i\lambda-\varepsilon)]
\label{rho4} ~.
\ee
The 
resolvent and the density in QCD$_3$ with an even number of flavors $2N_f$
follow analogously \cite{ADDV}.
In the replica formulation the resolvent is defined 
from the partition functions with $2N_f$ flavors with paired 
masses $\{\pm \mu_f\}$ as well as $2n$ replica flavors with
paired masses $\pm x$: 
\be 
G^{(2N_f)}_3(x,\{\mu_f\})=\lim_{n\rightarrow 0}
\frac{1}{2n}\frac{\partial}{\partial x}
\ln {\cal Z}_{3}^{(2N_f,2n)}(\{\mu_f\},x) ~.
\label{res3def}
\ee
As for the partition functions we use a similar notation as in QCD$_4$, the
two are easily separated by the explicit topological index $\nu$.  

\noi
The eigenvalue density is defined as in (\ref{rhodef}) except that the
average is in QCD$_3$ with $2N_f$ flavors. However, because the replicated
flavors have paired masses the discontinuity over the imaginary axis in the
complex quark mass plane of the resolvent becomes
\be
\frac{1}{2\pi} {\mbox{\rm Disc}}
\left.G_3^{(2N_f)}(x,\{\mu_f\})\right|_{x=i\lambda}
&=& \frac{1}{4\pi}\lim_{\varepsilon \to 0} 
\left\langle\sum_{k}\frac{2\varepsilon}
{(\lambda_k+\lambda)^2 +\varepsilon^2} + 
\frac{2\varepsilon}{(\lambda_k-\lambda)^2 +
\varepsilon^2}\right\rangle_{2N_f} \nonumber \\
&=& \frac{1}{2}\left(\rho_3^{(2N_f)}
(-\lambda,\{\mu_f\})+\rho_3^{(2N_f)}(\lambda,\{\mu_f\})\right) ~.
\label{rho3res}
\ee

\noi
The limit $n\to0$ in the defining equations for the resolvents (\ref{res4}) and
(\ref{res3def}) must obviously
be taken with care, since the partition functions entering the
right hand side are only known for integer values of $n$. For more than two
decades it was widely believed \cite{critique} that one could at best
obtain small or large argument expansions of the true result using the
replica method. 
Recent developments, see $e.g.$ \cite{KM,lerner}, attempted to go beyond this, 
but also that approach was met with some criticism \cite{zirn}.
However, with \cite{K,SplitVerb1,SplitVerb2,SplitVerb3} 
this situation has drastically changed. 
One tool to perform the replica limit correctly is the 
Toda lattice equations
for the leading-order QCD$_3$ partition functions. 
How it works in detail is perhaps best
explained by working out a couple of examples.

\noi
Let us start with the simplest case in QCD$_3$ namely the fully
quenched spectral density of the Dirac operator. 
In order to obtain this we first determine the fully quenched resolvent from
the Toda lattice equation (\ref{toda3})
\be
\lim_{n \to 0}\frac{1}{2n} (x\del_x)^2 \log {\cal Z}^{(2n)}_3(x) & = & x
\del_x x G_3^{(0)}(x) \nonumber \\ 
&=& 2x \left({\cal Z}^{(1)}_{3^+}(x){\cal Z}^{(-1)}_{3^-}(x)+
{\cal Z}^{(1)}_{3^-}(x){\cal Z}^{(-1)}_{3^+}(x)\right) ,
\label{repltoda3}
\ee
where we have used that ${\cal Z}^{(N_f=0)}_{3}=1$.
The partition functions can conveniently be found from (\ref{factorodd})
which then gives
\beq
\del_x x G_3^{(0)}(x) ~=~ x K_{1/2}(x)(I_{-1/2}(x)+I_{1/2}(x))
= 1 ~. \label{dG}
\eeq
The solution with boundary condition $G_3^{(0)}(x)$ finite in the limit
$x \to 0$ follows readily,   
\beq
G_3^{(0)}(x) = 1 ~.
\label{quenres3}
\eeq
Taking the discontinuity of the resolvent we get
\beq
\frac{1}{2}\left(\rho_3^{(0)}(-\lambda)+\rho_3^{(0)}(\lambda)\right) ~=~
\rho_3^{(0)}(\lambda) ~=~ \frac{1}{\pi},
\label{quenden3}
\eeq
where we have used that the even-flavored quenched spectral density
is an even function in $\lambda$. Eq. (\ref{quenden3}) 
is just the result derived from Random
Matrix Theory \cite{VZ}, where, due to the absence of the determinant,
it simply corresponds to the bulk microscopic UE density, which is flat.

\noi
Next, we work out the eigenvalue density in QCD$_3$ with an even
number of flavors. Using the same technical steps as in
\cite{SplitVerb1} the resolvent in QCD$_3$ with an
even number of flavors follows from the Toda lattice equation
(\ref{toda32}). Consider the easiest example $N_f=1$ which gives
\beq
\lim_{n\rightarrow 0}
\frac{1}{2n} x\partial_x (x\partial_x+y\partial_y)
\ln {\cal Z}_{3}^{(2n,2)}(x,y) ~=~ 2 x \frac{{\cal Z}^{(1,2)}_{3^+}(x,y) 
{\cal Z}^{(2,-1)}_{3^-}(y,x) + {\cal Z}^{(1,2)}_{3^-}(x,y) 
{\cal Z}^{(2,-1)}_{3^+}(y,x)}{\left[{\cal Z}^{(2)}_{3}(y)\right]^2} ~,
\label{repltoda32} 
\eeq
where once again the partition functions on the right hand side can be
found from (\ref{factoreven}) and (\ref{factorodd}). This can be
integrated to give a resolvent
\be
G^{(2)}_3(x,y) ~=~ 
\frac{x^2-y^2-x+ye^{-x}(\coth(y)\sinh(x)+\cosh(x)\tanh(y))}{x^2-y^2}
\label{pqres32}
\ee
and from this the density in QCD$_3$ with two paired masses follows:
\be
\rho^{(2)}_{3}(\lambda,y) ~=~ \rho^{(2)}_{3}(-\lambda,y) ~=~
\frac{1}{\pi}-\frac{y(\cos^2 \lambda \tanh y + \sin^2 \lambda \coth y)}
{\pi(\lambda^2+y^2)} ~.
\label{quenden32}
\ee
This is in complete agreement with the result found in Random Matrix Theory
\cite{DN}.

\noi
Now we move on to the odd sector. We look at (\ref{toda32unev}) with one
real flavor and $2n$ replicas for both an even and odd
number of eigenvalues. Taking the replica limit of this Toda lattice equation
gives us 
\beq
\lim_{n\rightarrow 0}
\frac{1}{2n} x\partial_x (x\partial_x+y\partial_y)
\ln {\cal Z}_{3^{\pm}}^{(2n,1)}(x,y) ~=~ 2 x \left(\frac{{\cal
Z}^{(1,2)}_{3^{\pm}}(x,y)  
{\cal Z}^{(2,-1)}_{3^{\mp}}(y,x)}{\left[{\cal Z}^{(2)}_{3}(y)\right]^2}
+ {\cal Z}^{(1)}_{3^{\mp}}(x)  
{\cal Z}^{(-1)}_{3^{\pm}}(x)\right) ~.
\label{repltoda32unev} 
\eeq
We start by looking at the case of an even number of eigenvalues. In
this case we get the following resolvent
\be
G^{(1)}_{3^+}(x,y) &=& 
\frac{x^2-y^2+e^{-x}(y\cosh(x)\tanh(y)-x\sinh(x))}{x^2-y^2}~,
\label{pqres32uep}
\ee
which then gives the symmetric part of the eigenvalue density in QCD$_3$ with
one flavor and an even number of eigenvalues  
\be
\frac{1}{2}\left(\rho^{(1)}_{3^+}(-\lambda,y)+
\rho^{(1)}_{3^+}(\lambda,y)\right) ~=~
\frac{1}{\pi}-\frac{\lambda \cos \lambda \sin \lambda +y \cos^2
\lambda \tanh y}{\pi (\lambda^2+y^2)} ~.
\label{quenden32uep}
\ee
In the massless case, $y=0$, this agrees with the prediction from Random
Matrix Theory \cite{C} and thus gives indirect confirmation of the 
conjectured form of the supersymmetric partition functions in QCD$_3$. 

\begin{figure}[!ht]
  \unitlength1.0cm
  \begin{center}
  \begin{picture}(3.0,1.3)
  \put(-5.,-5.){
  \psfig{file=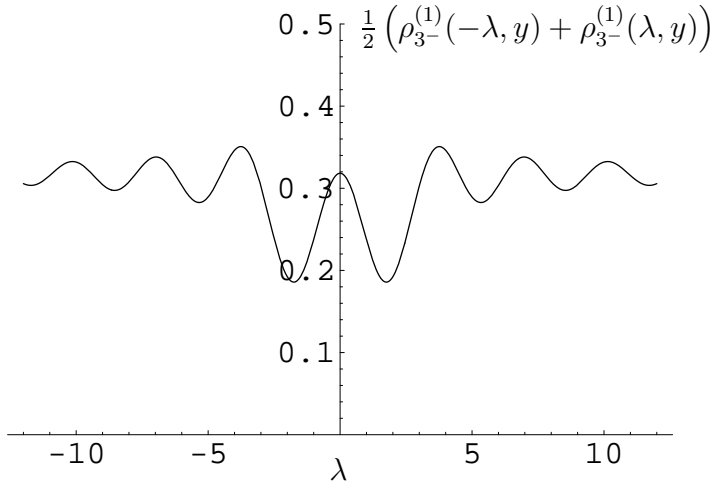,clip=,width=10cm}}
  \put(0,-5.1){\bf\large $\lambda$}
  \put(0.4,0.8){\bf\large $\frac{1}{2}\left(\rho^{(1)}_{3^-}
(-\lambda,y)+\rho^{(1)}_{3^-}(\lambda,y)\right)$}
  \end{picture}
  \vspace{4.5cm}
  \end{center}
\caption{\label{fig:oddodd1}Plot of the symmetric part of the
microscopic spectral density in QCD$_3$ for one massless quark and an odd
number of Dirac eigenvalues. The spectral density takes the value $1/\pi$ at
$\lambda=0$. }
\end{figure}

\begin{figure}[!ht]
  \unitlength1.0cm
  \begin{center}
  \begin{picture}(3.0,1.)
  \put(-5.,-5.){
  \psfig{file=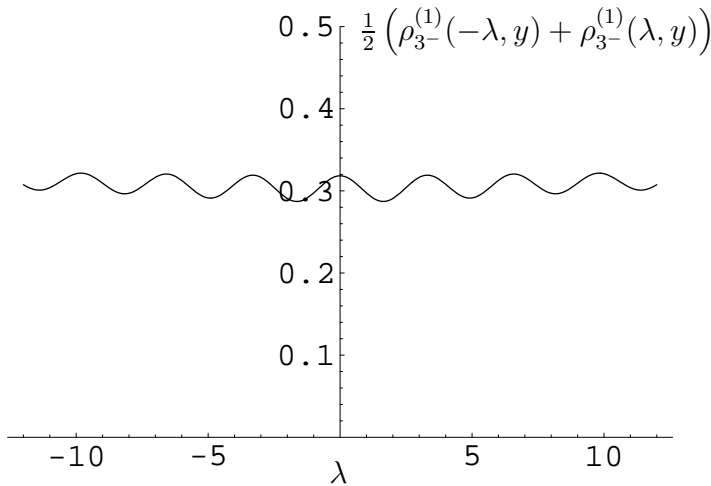,clip=,width=10cm}}
  \put(0,-5.1){\bf\large $\lambda$}
  \put(0.4,0.8){\bf\large $\frac{1}{2}\left(\rho^{(1)}_{3^-}
(-\lambda,y)+\rho^{(1)}_{3^-}(\lambda,y)\right)$}
  \end{picture}
  \vspace{4.5cm}
  \end{center}
\caption{\label{fig:oddodd2}Plot of the symmetric part of the spectral
  density for one  
relatively heavy quark, $y=10$, and an odd number of Dirac eigenvalues. 
The amplitude of the oscillations is smaller than in the
massless case. As the mass, $y$, goes to infinity the amplitude decrease 
as $1/y$, and the density converges
toward the quenched spectral density, $1/\pi$, as expected.}  
\end{figure}

\noi
We next consider the case of an odd number of Dirac operator eigenvalues,
for which microscopic spectral
correlation functions have not been evaluated previously due
to the unusual behavior of the partition function. 
This behavior does not obstruct our replica approach, 
in fact the evaluation of the spectral density is completely 
analogous to that in the even sector. The replica limit of the Toda lattice 
equation in this case gives the resolvent
\be 
G^{(1)}_{3^-}(x,y) &=&
\frac{x^2-y^2+e^{-x}(y\sinh(x)\coth(y)-x\cosh(x))}{x^2-y^2} ~,
\label{pqres32uem}
\ee
which then leads to the  even part of the density in QCD$_3$ with
one flavor and an odd number of eigenvalues
\be
\frac{1}{2}\left(\rho^{(1)}_{3^-}
(-\lambda,y)+\rho^{(1)}_{3^-}(\lambda,y)\right) ~=~
\frac{1}{\pi}+\frac{\lambda \cos \lambda \sin \lambda -y \sin^2
\lambda \coth y}{\pi (\lambda^2+y^2)} ~.
\label{quenden32uem}
\ee
This new result is plotted in 
figures \ref{fig:oddodd1} and \ref{fig:oddodd2}, the
plots showing respectively the massless case ($y=0$) and a 
massive case ($y=10$). 
Note in particular that in the massless case the sum does not vanish: 
the eigenvalue density in the massless case is {\sl not} an odd
function of $\lambda$. Let us clarify this point. It is tempting to assume 
that the spectral density of the Dirac operator eigenvalues will be
odd in $\lambda$. Our explicit calculation above shows that
it is not the case, but one can gain a better understanding of this
phenomenon if one is willing to use the Random Matrix Theory representation
for the eigenvalue density \cite{VZ}.
To this end, let $Z$ denote the partition function in the Random Matrix
Theory (only in the microscopic limit does this theory agree with the
static field integral \cite{VZ}), 
\be
Z^{(N_f)}_{3,N}(\{m_f\}) ~=~ \int \prod_{k=1}^N d\lambda_k 
\prod_{k<l=1}^N |\lambda_l-\lambda_k|^2 
\prod_{k=1}^Ne^{-NV(\lambda_k^2)}\prod_{f=1}^{N_f} (\lambda_k+ i m_f) 
\label{RMTZ}
\ee
for a suitable potential $V(\lambda^2)$ \cite{ADMN}.
We will use this representation to highlight the special 
properties of the density in
the theory with both $N_f$ and $N$ odd. It is sufficient to consider $N_f=1$. 
First note that the partition function
in this theory goes to zero linearly with $m$. This in itself is not alarming,
and for example also 
the effective QCD$_4$ partition function (\ref{ZchUEexpl}) 
vanishes like $m^{\nu}$ 
in the limit $m \to 0$. However in that case the analogous
Random Matrix Theory includes an explicit factor $m^{\nu}$ in front
of the eigenvalue representation, and this is easily canceled out when 
calculating the spectral density from the matrix integral. The small-mass
behavior of the integral (\ref{RMTZ}) is more complicated, as the mass $m$
mixes with the eigenvalues. To be precise, we simply {\em define} the
spectral density by the expectation value (\ref{rhodef}). From eq. 
(\ref{RMTZ}) it then follows that the eigenvalue density is given by
\be
\rho^{(1)}_{3,N}(\lambda_1,m)=\frac{(\lambda_1+ i m)e^{-NV(\lambda_1^2)}
\int \prod_{k=2}^N d\lambda_k 
\prod_{k<l=1}^N |\lambda_l-\lambda_k|^2 
\prod_{k=2}^Ne^{-NV(\lambda_k^2)} (\lambda_k+ i m)}{Z^{(1)}_{3,N}(m)} ~.
\ee
For odd $N$ and $\lambda_1\neq0$ the eigenvalue density diverges like $1/m$
for $m\to0$. However, for $\lambda_1=0$ the $m\to0$
limit is finite. Using the eigenvalue representation above it is also easy
to show that 
\be
\rho^{(1)}_{3,N}(-\lambda,m) = \rho^{(1)}_{3,N}(\lambda,-m)
\ee
and that the sum 
$$
\rho^{(1)}_{3,N}(\lambda,m) + \rho^{(1)}_{3,N}(-\lambda,m)
$$
appearing in (\ref{quenden32uem}), is finite for all values of $\lambda$
even in the limit $m\to0$. The divergence as $m \to 0$ thus resides entirely
in the odd part of the eigenvalue density, and this odd part is not probed by
the discontinuity of the resolvent.   
 
\noi
Our results in the odd-$N_f$ theory with an odd number of Dirac operator
eigenvalues therefore seem to be well understood also from the
Random Matrix Theory representation. As one additional non-trivial
check on our results, we note that the eigenvalue density (\ref{quenden32uem})
satisfies the correct decoupling condition of approaching the
fully quenched spectral density when the mass $y$ goes to infinity.




\vspace{1cm}

\noi
So far we have given several examples of how the eigenvalue density can be
derived from the replica limit of the Toda lattice equation. One can also
derive the chiral susceptibility in this fashion. This is in fact much
simpler since the Toda lattice equation already has a double derivative,
and no integration is therefore required. 

\noi
To determine the chiral susceptibility 
\beq
\chi_3^{(0)}(x,y) ~=~ \lim_{n\to 0, m\to 0}
\frac{1}{4nm}\frac{\partial}{\partial x} \frac{\partial}{\partial y}
\ln {\cal Z}^{(2m,2n)}_{3}(x,y) ~,
\eeq
we take the replica limit of the 
Toda lattice equation, (\ref{toda33}), and find
\beq
\chi_3^{(0)}(x,y) ~=~ 4 x^2  \left( \frac{{\cal
Z}_{3^-}^{(2,1)}(x,y) {\cal Z}_{3^+}^{(-2,-1)}(x,y)}{{\cal
Z}_{3^+}^{(1)}(x) {\cal Z}_{3^-}^{(-1)}(x)} + \frac{{\cal
Z}_{3^+}^{(2,1)}(x,y) {\cal Z}_{3^-}^{(-2,-1)}(x,y)}{{\cal
Z}_{3^-}^{(1)}(x) {\cal Z}_{3^+}^{(-1)}(x)} \right) ~.
\eeq
Using Theorem II and (\ref{bosth2}) we find 
\beq
\chi_3^{(0)}(x,y) ~=~ 
\frac{e^{-x-y} \sinh (x+y)}{(x+y)^2}~. \label{chirep}
\eeq
The double discontinuity of this function gives the double symmetric
combination of the two point function \cite{ADDV}
\be
 \rho_3^{(0)}(\lambda_1,\lambda_2)+\rho_3^{(0)}(-\lambda_1,\lambda_2)
+\rho_3^{(0)}(\lambda_1,-\lambda_2)+\rho_3^{(0)}(-\lambda_1,-\lambda_2) =
\frac{1}{(2\pi)^2} {\mbox{\rm Disc}}
\left.\chi_3^{(0)}(x,y)\right|_{x=i\lambda_1,y=i\lambda_2} ~.
\ee

\noi
We note that all of these cases, and any other we can consider, give
the detailed support for the general relations that formally can be
derived on the basis of Theorems I and II and the replica definitions
(\ref{res4}) and (\ref{res3def}) alone \cite{ADDV}. These are, for the
one-point functions,
\be 
\rho_{3}^{(2N_{f})}(\lambda;\{\mu_i\})&=&
\frac{1}{2}\left[\rho_{{\rm QCD}_4}^{(N_{f},\nu=+1/2)}(\lambda;\{\mu_i\}) +
\rho_{{\rm QCD}_4}^{(N_{f},\nu=-1/2)}(\lambda;\{\mu_i\})\right]
\label{rho34even}
\ee
for an even number of flavors, and 
\be
\rho_{3^+}^{(2N_{f}+1)}(\lambda;\{\mu_i\},\mu) +
\rho_{3^+}^{(2N_{f}+1)}(-\lambda;\{\mu_i\},\mu)  ~=~ 
\rho_{\nu=-1/2}^{(N_{f}+1)}(\lambda;\{\mu_i\},\mu) +
\rho_{\nu=+1/2}^{(N_{f})}(\lambda;\{\mu_i\}) ~,
\label{rho34odd+}
\ee
for an odd number of flavors in the ``+''-sector.\footnote{Note that
ref. \cite{ADDV} incorrectly assumed that $\rho_{3^+}^{(2N_{f}+1)}
(\lambda;\{\mu_i\},\mu)$ is an even function of $\lambda$ also in the
case of $\mu \neq 0$.}

\noi
In ref. \cite{ADDV} the case of an odd number of flavors in the
``-''-sector was not considered, but we can easily derive the analogous
relation by use of Theorem II. We find
\be
\rho_{3^-}^{(2N_{f}+1)}(\lambda;\{\mu_i\},\mu) +
\rho_{3^-}^{(2N_{f}+1)}(-\lambda;\{\mu_i\},\mu)  ~=~ 
\rho_{\nu=+1/2}^{(N_{f}+1)}(\lambda;\{\mu_i\},\mu) +
\rho_{\nu=-1/2}^{(N_{f})}(\lambda;\{\mu_i\}) ~,
\label{rho34odd-}
\ee
for which our (\ref{quenden32uem}) indeed is a special case.

\noi
We also remind that similar general relations can be worked out for
higher $k$-point spectral correlation functions \cite{ADDV}, and that
Toda lattice equations can give the justification for the formal
replica manipulations that lead to these relations. Because of the
pairing of masses with different signs these relations do not just involve
the $k$-point spectral correlators themselves, but combinations with
different signs (see ref. \cite{ADDV} for details). We have already
seen examples of this phenomenon for the spectral 1-point functions
derived above. Also
the chiral susceptibility (\ref{chirep}) gives the symmetric combination of
the spectral 2-point function. To isolate
the spectral correlation functions directly one needs to combine
the replica method with more than just the discontinuity formula
of the mass-paired resolvent and higher-order versions thereof.

\vspace{0.5cm}
\section{Graded Toda lattice equation}
\label{sec:susy}
\vspace*{0.3cm}

While the Toda lattice equation allows us to take the replica limit in a well
defined and correct way it does not fully explain why the analytic
continuation in the number of flavors works. However, it is not difficult to
understand why the replica limit of the Toda lattice equation and the 
supersymmetric \cite{Efetov} method give the same answer. To make this
transparent \cite{SplitVerb3} we
consider supersymmetric versions of the Toda lattice equations and quench
these as in the supersymmetric method \cite{Efetov}. Referring to the graded
symmetry such Toda lattice equations have been called 
graded Toda lattice equations \cite{SplitVerb3}. 

\noi
Two graded Toda lattice equations have been found for
QCD$_4$ \cite{SplitVerb3}. Here we will take a closer look at one of these
namely the supersymmetric variant of (\ref{toda42}) with $N_f$ degenerate
fermionic flavors of paired masses $\pm x$ and $N_b$ degenerate
bosonic flavors of paired masses $\pm y$
\beq
x\partial_x(x\partial_x +y\partial_y)\ln{\cal Z}^{(N_f|N_b)}_{\nu}(x|y) ~=~
2N_fx^2\frac{{\cal Z}^{(N_f+1|N_b)}_{\nu}(x|y)
{\cal Z}^{(N_f-1|N_b)}_{\nu}(x|y)}
{\left[{\cal Z}^{(N_f|N_b)}_{\nu}(x|y)\right]^2} ~. \label{susytoda42}
\eeq
The generalized versions of Theorems I and II can be used to translate this
equation in to a graded Toda lattice equation for QCD$_3$
\be
&& x\partial_x (x\partial_x + y\partial_y) 
\ln {\cal Z}^{(2N_f|2N_b)}_{3}(x|y) \nonumber \\ 
&=& 4 N_f x \frac{{\cal Z}^{(2N_f+1|2N_b)}_{3^+}(x|y) 
{\cal Z}^{(2N_f-1|2N_b)}_{3^-}(x|y) + {\cal Z}^{(2N_f+1|2N_b)}_{3^-}(x|y) 
{\cal Z}^{(2N_f-1|2N_b)}_{3^+}(x|y)}{\left[{\cal
Z}^{(2N_f|2N_b)}_{3}(x|y)\right]^2 
} ~. 
\label{susytoda32}
\ee
It suffices to
look at the special case of $N_f=1$ and $N_b=1$, and quench it
supersymmetrically. We start by focusing on the lhs  
\be
\lim_{y \to x} \frac{1}{2} x\del_x (x \del_x +y \del_y) \ln {\cal
Z}_{3}^{(2|2)}(x,-x|y,-y) 
&=& x \del_x x G_3^{(0)}(x) ~,
\label{susyres}
\ee
where we have used the supersymmetric 
definition of the fully quenched resolvent
\be 
G_3^{(0)}(x) \equiv \lim_{y\to x}\del_x\ln {\cal Z}_{3}^{(2|2)}(x,-x|y,-y)~.
\ee
Taking $y\to x$ on the rhs of (\ref{susytoda32}) with $N_f=N_b=1$ 
thus leads to
\beq
x \del_x x G_3^{(0)}(x) ~=~ 2 x \left({\cal Z}_{3^+}^{(1)}(x){\cal
Z}_{3^-}^{(-1)}(x) + {\cal Z}_{3^-}^{(1)}(x){\cal Z}_{3^+}^{(-1)}(x)
\right) ~,
\label{susyresquen}
\eeq
in complete consistency with (\ref{dG}) derived from the replica limit of the
Toda lattice equation. This immediate consistency comes about since the
supersymmetric generating functionals satisfy exactly the same Toda lattice
equation as the fermionic and bosonic hierarchy of partition functions.

\vspace{0.5cm}
\section{Consistency Conditions and Toda Lattice Equations}
\label{sec:consist}
\vspace*{0.3cm}

It is interesting to note that the Toda equations for QCD$_4$ are
special cases 
of a more general relation which was derived in ref. \cite{AD}, and there
referred to as ``Consistency Condition II'' due to this equation's
origin as a consistency condition in Random Matrix Theory \cite{AD1}.
This more general relation reads, after rotating to real masses,
\be
{\cal Z}_{\nu}^{(N_f+2)}(x,y,\{\mu_f\}) &=& 
\frac{1}{(x^2-y^2){\cal Z}_{\nu}^{(N_f)}(\{\mu_f\})} \label{consist4}\\
&\times& \left[{\cal Z}_{\nu}^{(N_f+1)}(y,\{\mu_f\})(\sum_{f=1}^{N_f} 
\mu_f \partial_{\mu_f} +x\partial_x) 
{\cal Z}_{\nu}^{(N_f+1)}(x,\{\mu_f \})
-(x\leftrightarrow y) \right]~.\nn
\ee
Taking the completely mass-degenerate case $x=y=\mu_f$ for all $f$, and
switching to the conventional normalization for the mass-degenerate case, 
we recover 
the Toda lattice eq. (\ref{toda3}). Similarly, if we take $x=y=\mu_f$ for
$f= 1,\ldots,N_f$ and $\mu_f=z$ for $f=N_f+1,\ldots,N_f+n$ we recover the
Toda eq. (\ref{toda42}). It appears that the single equation (\ref{consist4})
is the generator of all Toda lattice equations. 

\noi
In ref. \cite{AD1} it was stated that an analogous consistency condition
exists for the QCD$_3$ case, but no details were given. Here we supply this
missing equation, which turns out to be of a quite different kind. We begin
by noting that the kernel $K(x,y,\{\mu_f\})$ associated with the Random
Matrix Theory (\ref{RMTZ}) can be written in terms of the even-flavored
effective QCD$_3$ partition function \cite{AD1}
\beq
K(x,y,\{\mu_i\}) = \prod_f^{N_f}\sqrt{(x^2+\mu_f^2)(y^2+\mu_f^2)}
\frac{{\cal Z}_{3}^{(2N_f+2)}(ix,iy,\{\mu_i\})}
{{\cal Z}_{3}^{(2N_f)}(\{\mu_i\})} ~,
\label{kernelZ}
\eeq
while in general from Random Matrix Theory it is known that it also
can be represented in terms of the associated orthogonal polynomials
$P_n(x,\{\mu_i\})$ (see $e.g.$, ref. \cite{ADMN} for details),
\beq
K(x,y,\{\mu_i\}) = \prod_f^{N_f}\sqrt{(x^2+\mu_f^2)(y^2+\mu_f^2)}
\frac{P_{2N-1}(x,\{\mu_i\})P_{2N}(y,\{\mu_i\})-
P_{2N}(x,\{\mu_i\})P_{2N-1}(y,\{\mu_i\})}{x-y} ~.
\label{kernelP}
\eeq
For this particular Random Matrix Theory ensemble the orthogonal
polynomials split into two disjoint sectors of odd and even order.
Now, since also these orthogonal polynomials in the large-$N$ limit
can be expressed directly
in terms of the finite-volume partition functions \cite{AD1},
\begin{eqnarray}
P_{2N}(x,\{\mu_i\}) ~=~ \frac{{\cal Z}_{3^+}^{(2N_f+1)}(ix,\{\mu_i\})}
{{\cal Z}_{3}^{(2N_f)}(\{\mu_i\})} \cr
P_{2N+1}(x,\{\mu_i\}) ~=~ \frac{{\cal Z}_{3^-}^{(2N_f+1)}(ix,\{\mu_i\})}
{{\cal Z}_{3}^{(2N_f)}(\{\mu_i\})} \label{PZ}
\end{eqnarray}
we can combine eqs. (\ref{kernelZ})-(\ref{PZ}) into a consistency
condition that must be satisfied by the effective partition functions.
In contrast to the QCD$_4$ case \cite{AD1} we do not need to expand
to first non-trivial order in $1/N$, and the relation will therefore
be {\em algebraic}. We find, after rotating into real masses throughout
and fixing the normalization constant,
\begin{eqnarray}
&& {\cal Z}_{3}^{(2N_f+2)}(x,y,\{\mu_i\}) ~=~ \cr
&&\frac{2}{x-y}\frac{
{\cal Z}_{3^-}^{(2N_f+1)}(x,\{\mu_i\}){\cal Z}_{3^+}^{(2N_f+1)}(y,\{\mu_i\})
- {\cal Z}_{3^-}^{(2N_f+1)}(x,\{\mu_i\}){\cal Z}_{3^+}^{(2N_f+1)}(y,\{\mu_i\})}
{{\cal Z}_{3}^{(2N_f)}(\{\mu_i\})} ~.
\label{consist3}
\end{eqnarray}
The three main types of partition functions that enter QCD$_3$ are
thus not independent.

\noi
A particularly useful special case of eq. (\ref{consist3}) is that 
of $y=-x$. Using ${\cal Z}_{3^-}^{(2N_f+1)}(-x,\{\mu_i\}) =
- {\cal Z}_{3^-}^{(2N_f+1)}(x,\{\mu_i\})$ as well as
${\cal Z}_{3^+}^{(2N_f+1)}(-x,\{\mu_i\}) =
{\cal Z}_{3^+}^{(2N_f+1)}(x,\{\mu_i\})$ we get a simple factorization
identity,
\beq
{\cal Z}_{3}^{(2N_f+2)}(x,-x,\{\mu_i\}) ~=~
\frac{2}{x}\frac{{\cal Z}_{3^+}^{(2N_f+1)}(x,\{\mu_i\})
{\cal Z}_{3^-}^{(2N_f+1)}(x,\{\mu_i\})}{{\cal Z}_{3}^{(2N_f)}(\{\mu_i\})} ~.
\label{consist3simple}
\eeq
This special case actually also follows from combining the
relations of Theorems I-II, and re-expressing results in terms
of QCD$_3$ partition functions alone. However, the more general relation
(\ref{consist3}) seems to go beyond what is contained in these 
theorems.

\noi
The relation (\ref{consist3simple}) gives rise to non-trivial
identities among spectral correlation functions if we combine it
with the replica method in the same way we derived the general relations
(\ref{rho34even})-(\ref{rho34odd-}). We can illustrate this very easily by
focusing on the spectral density itself. To this end, consider
the special case where we add $2N$ fermions of paired (common up to a sign) 
masses $\pm\mu$
to the $2N_f$ paired masses $\mu_i$. 
The resolvents of the different theories can then be related:
\be
&&G_3^{(2N_f+2)}(x,-x,\{\mu_i\},\mu) = \lim_{N\to 0}\frac{1}{2N}
\partial_{\mu}\ln{\cal Z}_{3}^{(2N_f+2N+2)}(x,-x,\{\mu_i\},\mu) \cr
&=&  \lim_{N\to 0}\frac{1}{2N}\left[
\partial_{\mu}\ln{\cal Z}_{3^+}^{(2N_f+2N+1)}(x,\{\mu_i\},\mu)
+ 
\partial_{\mu}\ln{\cal Z}_{3^-}^{(2N_f+2N+1)}(x,\{\mu_i\},\mu)\right.
\cr && - \left.
\partial_{\mu}\ln{\cal Z}_{3}^{(2N_f+2N)}(\{\mu_i\},\mu)\right]\cr
&=& G_{3+}^{(2N_f+1)}(x,\{\mu_i\},\mu) + G_{3-}^{(2N_f+1)}(x,\{\mu_i\},\mu)
- G_3^{(2N_f)}(\{\mu_i\},\mu)
\ee  
which in turn implies an identity among the spectral densities,
\be
&& 2\left[\rho_{3}^{(2N_f+2)}(\lambda,x,-x,\{\mu_i\}) + 
\rho_{3}^{(2N_f)}(\lambda,\{\mu_i\})\right] = 
\rho_{3^+}^{(2N_f+1)}(\lambda,x,\{\mu_i\}) + 
\rho_{3^+}^{(2N_f+1)}(-\lambda,x,\{\mu_i\})\cr
&& +\rho_{3^-}^{(2N_f+1)}(\lambda,x,\{\mu_i\}) + 
\rho_{3^-}^{(2N_f+1)}(-\lambda,x,\{\mu_i\})  ~.
\ee
Needless to say, this general relation also follows from combining
eqs. (\ref{rho34even})-(\ref{rho34odd-}).

\section{Conclusions}

The purpose of this paper was to demonstrate that the replica method based on
Toda lattice equations successfully can be applied to QCD$_3$ in
the $\epsilon$-regime. The result is a series of exact statements
about spectral correlation functions for the Dirac operator of that 
theory in the phases where spontaneous symmetry breaking occurs. The
Toda lattice equations were derived on the basis of known Toda equations
for the effective QCD$_4$ partition functions in the $\epsilon$-regime
and exact relations between the two theories. Our results give the
detailed support for general relations that formally can be derived
by the replica method applied directly on the relations between
QCD$_3$ and QCD$_4$. Because our corresponding Random Matrix Theory
ensemble incorporate a number of determinants, these results go much
beyond, but include as a special case, what has already been shown
for the ordinary unitary ensemble.

\noi
Armed with a reliable tool for computing spectral correlation functions
by means of the replica method, we have also attacked a case that 
previously has not been considered in detail in the literature: an
odd number of flavors in a regularization with an odd number of
Dirac operator eigenvalues. We have shown that the peculiarities surrounding 
this case do not lead to pathologies in the spectral properties of
the theory. This can also be understood from the point of view
of the corresponding Random Matrix Theory once the singularity
at vanishing quark mass has been treated carefully.

\noi
As in 4-dimensional QCD, the replica method based on Toda lattice
equations allows for a smooth connection to the supersymmetric
approach. In particular we have conjectured the 
form of the supersymmetric generalization of the relationship between
the effective QCD$_3$ and QCD$_4$ partition functions in the
extreme limit of the $\epsilon$-regime. With the corresponding
4-dimensional expressions known in closed analytical form this
gives us a very compact expression for the analogous expression
in QCD$_3$, an expression which will be extremely tedious and 
difficult to derive directly in the chiral Lagrangian form (previous
results in that direction do not include the so-called Efetov-Wegner
``boundary'' terms \cite{S}). We
suspect that a simple proof may be obtained by reverting to
the Random Matrix Theory expression for the partition function.

\noi
Finally we have illustrated how Toda lattice equations for QCD$_4$
follow as special cases of a more general expression
derived originally from the Random Matrix Theory representation.
Interestingly, when we extend the proof in the most straightforward
manner to QCD$_3$ we do not recover our Toda lattice equations,
but instead an algebraic relationship between the involved effective
partition functions. Again, as a special case this relation can be
reduced to an equation that can also be derived on the basis of
the connection between QCD$_3$ and QCD$_4$. When applying the
replica method to this relation, now justified on account of
the Toda approach, we obtain exact relations between the
spectral correlation functions involved. We have verified explicitly 
in simple cases that this general constraint indeed is
satisfied.

\noi\noindent
{\bf Acknowledgments}\\ We thank Jac Verbaarschot for inspiring
discussions on 
the Toda lattice equations in QCD and Gernot Akemann for helpful
discussions on the consistency condition. We also wish to thank 
Thomas Seligman, Thomas Guhr and Luis Benet for hospitality at 
Centro Internacional de Ciencia in Cuernavaca where part of this work was
completed.

\vspace{0.6cm}
\appendix
\setcounter{equation}{0}

\end{document}